\documentclass[%
 aip,
 jmp,%
 amsmath,amssymb,
reprint,%
]{revtex4-1}

\usepackage{graphicx}
\usepackage{dcolumn}
\usepackage{bm}
\usepackage{subfigure} 

\begin{document}


\title[Parallel and series nucleation]{A note on the nucleation with multiple steps: \\Parallel and series nucleation
}

\author{Masao Iwamatsu}
\email{iwamatsu@ph.ns.tcu.ac.jp}
\affiliation{ 
Department of Physics, Tokyo City University, Setagaya-ku, Tokyo 158-8557, JAPAN
}%


\date{\today}

\begin{abstract}
Parallel and series nucleation are the basic elements of the complex nucleation process when two saddle points exist on the free-energy landscape.  It is pointed out that the nucleation rates follow formulas similar to those of parallel and series connection of resistors or conductors in an electric circuit.  Necessary formulas to calculate individual nucleation rates at the saddle points and the total nucleation rate are summarized and the extension to the more complex nucleation process is suggested.
\end{abstract}

\maketitle

\section{\label{sec:level1x}Introduction}
Nucleation is a very basic phenomenon which plays a vital role in various materials processing ranging from steel production to food beverage industries~\cite{Kelton2010}.  Recently, researchers' interests have turned to complex material nucleations which are relevant to our daily life~\cite{Kelton2010}.  The nucleation of such complex materials can also be complex and may consist of multiple steps~\cite{Vekilov2004, Gebauer2008}.  These multiple steps correspond to the multiple saddle points on the free energy landscape and can occur in  series or in parallel.  Therefore, the nucleation with multiple steps can be akin to the text-book problem of electric circuitry. 

The most basic unit of electric circuitry is the parallel or series connections of two resistors.  Likewise, there are some problems of nucleation where there are two saddles points on the free-energy landscape.  For example, Ray et al.~\cite{Ray1986} found the two saddle points (double barriers) on the two-dimensional free-energy landscape of a droplet condensation of binary vapor~\cite{Ray1986}. A sequential two saddle points on the free-energy landscape along a fictitious one-dimensional reaction coordinate has been studied by Valencia and Lipowsky~\cite{Valencia2004} to study the condensation on a patterned substrate.  These free-energy landscapes imply two successive {\it series} nucleations. A similar double-barrier model was assumed to study the nucleation and growth by various authors~\cite{Vekilov2004,Kashchiev1998,Kashchiev2005} when an intermediate metastable phase exists.  

\begin{figure}[htbp]
\begin{center}
\includegraphics[width=0.65\linewidth]{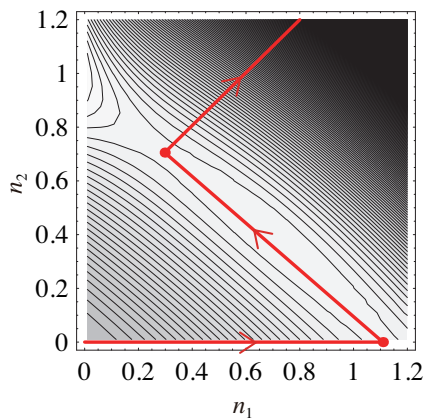}
\end{center}
\caption{
An example of the free-energy landscape of composite nucleus that consists of a stable solid ($n_2$) surrounded by an intermediate metastable liquid ($n_1$).  This is calculated using the capillarity approximation~\cite{Iwamatsu2011} in the (metastable liquid ($n_1$), stable solid ($n_2$))-space. The two red points indicate two saddle points with the same free-energy barriers which correspond to the critical nucleus of metastable liquid and the critical composite nucleus composed from solid core and surrounding metastable liquid wetting layer.  The red line indicates the series nucleation route suggested by previous authors~\cite{Vekilov2004,Kashchiev1998,Kashchiev2005}. This free-energy landscape with two saddle points clearly indicates a parallel nucleation (two nucleation routes). } 
\label{fig:1x}
\end{figure}

Recently, the present author~\cite{Iwamatsu2011} re-examined the free-energy landscape using capillarity approximation when an intermediate metastable phase existed.  The location of the two saddle points on the free-energy landscape seems to suggest a {\it parallel} nucleation in contrast to the assumption of a {\it series} nucleation of previous authors~\cite{Vekilov2004,Kashchiev1998,Kashchiev2005}.  Figure~\ref{fig:1x} shows the free-energy landscape when two saddle points have the same free-energy barrier~\cite{Iwamatsu2011}.  The two red points indicate two saddle points and the red line indicates the series nucleation route assumed by previous authors~\cite{Vekilov2004,Kashchiev1998,Kashchiev2005}.  This figure clearly suggests a parallel nucleation with two nucleation routes which pass through two different saddle points.  Similar free-energy landscapes with two types of critical nucleus of different compositions have been predicted by Ray et al.~\cite{Ray1986} for partially miscible binary systems such as diethylene glycol/{\it n}-heptane, methanol/cyclohexane and diethlene glycol/benzen  from the free-energy calculation using the capillarity approximation, and by Chen et al.~\cite{Chen2003} for water/{\it n}-nonane mixtures using Monte Carlo algorithm with the umbrella sampling technique.  Experimental result for the water/{\it n}-nonane mixture by Wagner and Strey~\cite{Wagner2001} is also consistent to the picture of parallel nucleation with two saddle points.  Therefore, {\it parallel} and {\it series} nucleation seem to be common basic units of various complex nucleation phenomena.

In this report, we consider the most basic elements of complex nucleation.  A series and a parallel nucleation, which are akin to the basic units of electric circuit: series and parallel connection of resistors or conductors.  We summarize the basic ingredients of the parallel and the series nucleations and emphasize the complex nucleation process which could be sorted out by mapping the complex nucleation process into a simple electric circuitry.

\section{\label{sec:level2x}Continuum description of nucleation process}
Nucleation kinetics can be studied using the discrete or continuum descriptions~\cite{Kelton2010,Wu1997}.  In this report, we use the continuum description since we want to make correspondence to the classical electrostatics more transparent.  In the continuum description, the discrete Master equation is transformed into the Fokker-Planck equation~\cite{Wu1997, Risken1989} or so-called Zeldovich-Frenkel equation~\cite{Zeldovich1942,Frenkel1955} for the one-component system.

We start from the general Fokker-Planck equation for the multi-component system~\cite{Reiss1950} of the form
\begin{equation}
-\frac{\partial f\left({\bm n}\right)}{\partial t}=-\nabla\cdot{\bm J}
\label{eq:1x}
\end{equation}
where
\begin{equation}
f\left({\bm n}\right)=f\left(n_1, n_2, \dots\right)
\label{eq:2x}
\end{equation}
is the number of clusters with composition ${\bm n}=\left(n_1, n_2, \dots\right)$, and ${\bm J}$ is the nucleation flux vector
\begin{equation}
{\bm J}=J_1{\bm e}_1+J_2{\bm e}_2+\cdots.
\label{eq:3x}
\end{equation}
where ${\bm e}_i$ are unit vectors in the Cartesian $(n_1, n_2, \dots)$ coordinate.  By assuming the detailed balance at the equilibrium~\cite{Reiss1950}, the flux components $J_i$ are derived from the potential $\Phi$ defined by
\begin{equation}
\Phi\left(n_1, n_2, \dots\right)=\frac{f\left(n_1, n_2, \dots\right)}{N\left(n_1, n_2, \dots\right)},
\label{eq:4x}
\end{equation}
where
\begin{equation}
N\left(n_1, n_2, \dots\right)=N_0\exp\left(-\frac{W\left(n_1, n_2, \dots\right)}{kT}\right)
\label{eq:5x}
\end{equation}
is the equilibrium number of clusters and $W$ is the reversible work (free-energy) of cluster formation with composition $n_1, n_2, \dots$.

Using the potential $\Phi$, we can write the nucleation flux as
\begin{equation}
J_i = -N\left(n_1, n_2, \dots\right)R_i\left(n_1, n_2, \dots\right)\frac{\partial \Phi}{\partial n_i}
\label{eq:6x}
\end{equation}
where $R_i$ represents the reaction rates of nucleation which will take different forms in vapor condensation~\cite{Reiss1950} and nucleation in condensed phase~\cite{Temkin1984}.  Equations (\ref{eq:1x}) and (\ref{eq:6x}) lead to the Fokker-Planck equation~\cite{Risken1989} or Frenkel-Zeldovich~\cite{Frenkel1955,Zeldovich1942} equation when we consider only one component.  Therefore, we have to investigate the Brownian motion on a multi-dimensional surface of the potential $\Phi$.  

Since we are interested in the series and parallel nucleation phenomena, which are the most basic elements of complex nucleation, we consider the nucleation by a two-dimensional potential $\Phi\left(n_1, n_2\right)$, or the binary nucleation of two-component system that has been considered thoroughly~\cite{Stauffer1976,Trinkaus1983,Greer1990,Wilemski1995,Wyslouzil1995,Wilemski1999,Fisenko2004}.  We concentrate on the steady-state nucleation rate only though previous authors~\cite{Stauffer1976,Trinkaus1983,Greer1990,Wilemski1995,Wyslouzil1995,Wilemski1999,Fisenko2004} were interested mostly in the relations between the nucleation current, the steepest-descent direction of the free energy landscape (work of formation) and the gradient direction of the potential $\Phi$.

\subsection{\label{sec:level2}Parallel nucleation}
In the two-dimensional space $(n_1,n_2)$, we have to solve the Fokker-Planck equation:
\begin{equation}
-\frac{\partial f\left({\bm n}\right)}{\partial t}=\nabla\cdot{\bm J}=\frac{\partial J_1}{\partial n_1}+\frac{\partial J_2}{\partial n_2}
\label{eq:7x}
\end{equation}
subject to the boundary condition~\cite{McDonald1963}
\begin{equation}
f\left(\left|{\bm n}\right| \rightarrow \infty \right)=0,\;\;\;f\left(\left|{\bm n}\right| \rightarrow 0 \right)=N\left({\bm n}\right).
\label{eq:8x}
\end{equation}
The steady-state nucleation flux ${\bm J}^{\rm st}$ can be studied when $\partial f/\partial t=0$, which leads to
\begin{equation}
\frac{\partial J_1^{\rm st}}{\partial n_1}+\frac{\partial J_2^{\rm st}}{\partial n_2}=0
\label{eq:9x}
\end{equation}
from Eq.~(\ref{eq:7x}).  Since $\mbox{div} {\bm J}^{\rm st}=0$, in arbitrary closed region on the two-dimensional space $(n_1,n_2)$, the flux must be conserved.  In other words, the flux coming into this region must be balanced to the flux going out from this region. 

\begin{figure}[htbp]
\begin{center}
\subfigure[A scenario of parallel nucleation]
{
\includegraphics[width=0.55\linewidth]{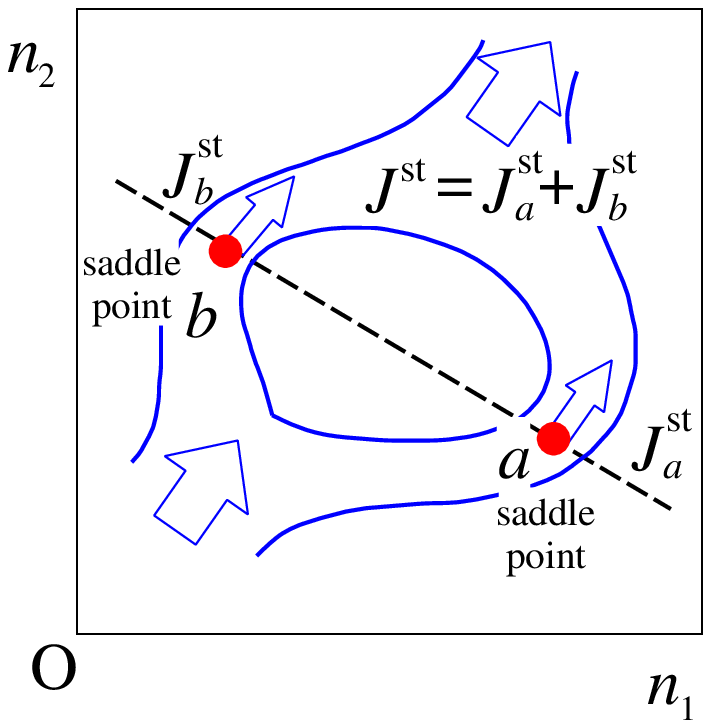}
\label{fig:2xa}
}
\subfigure[A scenario of series nucleation]
{
\includegraphics[width=0.55\linewidth]{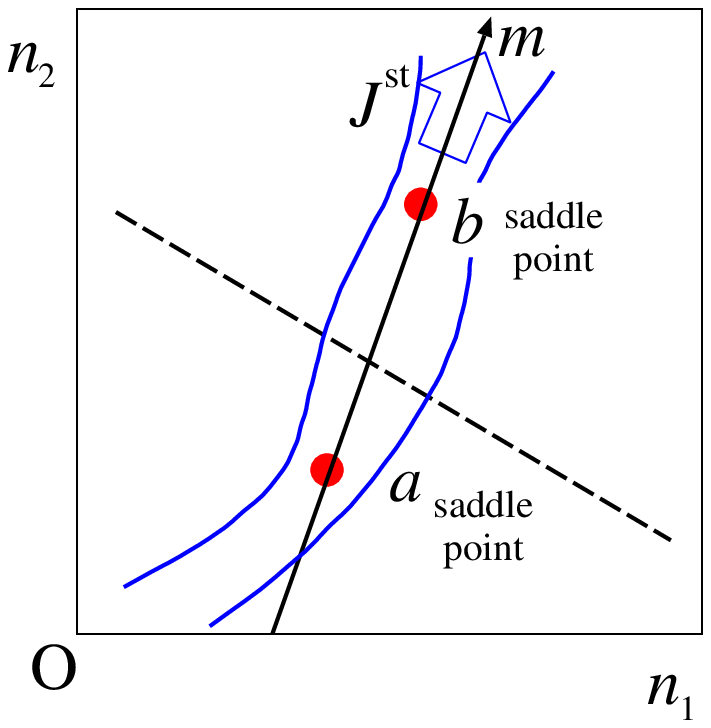}
\label{fig:2xb}
}
\end{center}
\caption{
A typical scenario of (a) parallel and (b) series nucleation on a free-energy landscape of the simplest two-component system:  The nucleation flux ${\bm J}^{\rm st}$ indicated by the wide arrow must be continuous and conserved as there is no source term in Eqs.~(\ref{eq:9x}) and (\ref{eq:11x}).  In (b), the problem is reduced to the one-dimensional problem along the fictitious coordinate $m$.
 } 
\label{fig:2x}
\end{figure}

It is customary to assume the nucleation flux flow through the saddle point of the free-energy landscape $W\left(n_1,n_2\right)$~\cite{Stauffer1976,Trinkaus1983,Greer1990,Wilemski1995,Wyslouzil1995,Wilemski1999,Fisenko2004}.  Suppose we have two saddle points, say $a$ and $b$ whose positions are almost in parallel along the line $n_1+n_2\sim\mbox{constant}$.  Therefore, the two saddle points correspond to the two types of nucleus of almost the same size and with different composition as shown in Fig.~\ref{fig:2x}(a).  In such a case, it is reasonable to assume the nucleation flux ${\bm J}^{\rm st}$ will fork into two branches ${\bm J}_{a}^{\rm st}$ that passes through the saddle point $a$ and ${\bm J}_b^{\rm st}$ that passes trough $b$ as shown schematically in Fig.~\ref{fig:2x}(a), and the flux far from the two saddle points becomes the sum of the two fluxes
\begin{equation}
{\bm J}^{\rm st} = {\bm J}_a^{\rm st}+{\bm J}_b^{\rm st},
\label{eq:10x}
\end{equation}
where these two fluxes satisfy the conservation law
\begin{equation}
\mbox{div}{\bm J}_{\kappa}^{\rm st}=0,\;\;\;\kappa=a,b
\label{eq:11x}
\end{equation}
independently around the two saddle points $a$ and $b$.  This simple equation (\ref{eq:10x}) which represents the superposition of two independent nucleations has already been suggested and used by Wagner and Strey~\cite{Wagner2001} to explain their experimental result of nucleation in water/{\it n}-nonane vapor mixture.

General solutions for the steady-state nucleation rate at the saddle point were given by Trinkaus~\cite{Trinkaus1983}.  First of all, we expand the work of formation $W\left({\bm n}\right)$ around the saddle points ${\bm n}_{\kappa}^{*}=\left(n_{\kappa, 1}^{*},n_{\kappa, 2}^{*}\right), (\kappa=a, b)$:
\begin{eqnarray}
W\left(n_1,n_2\right) &\simeq& W_{\kappa}^{*} + \frac{1}{2}\sum_{i,j}\left(n_i-n_{\kappa, i}^{*}\right) W_{\kappa, ij}^{*} \left(n_j-n_{\kappa, j}^{*}\right), \nonumber \\
&&\qquad\qquad\qquad\qquad\qquad\qquad\kappa=a,b
\label{eq:12x}
\end{eqnarray}
where
\begin{equation}
W_{\kappa}^{*}=W\left(n_{\kappa, 1}^*,n_{\kappa, 2}^*\right)
\label{eq:13x}
\end{equation}
is the free-energy barrier at the saddle point $\kappa=a,b$ and
\begin{equation}
W_{\kappa, ij}^{*}=\left(\frac{\partial^2 W}{\partial n_i \partial n_j}\right)_{\bm{n}=\bm{n}_{\kappa}^{*}}.
\label{eq:14x}
\end{equation}
However, the steepest-descent direction of neither the free-energy landscape~\cite{Reiss1950} $W$ nor the potential landscape $\Phi$ does not necessarily tell you the direction of nucleation flux since the reaction rate $R_i$ can be different~\cite{Stauffer1976} $R_1\neq R_2$.

In order to take into account the anisotropy of $R$, according to Trinkaus~\cite{Trinkaus1983} and Wilemski~\cite{Wilemski1999}, we introduce new variables ${\bm \nu}_\kappa=(\nu_{\kappa, 1},\nu_{\kappa, 2})$ through
\begin{eqnarray}
n_1-n_{\kappa, 1}^*=\left(R_{\kappa,1}^{*}\right)^{1/2}\nu_{\kappa, 1},
\label{eq:15x} \\
n_2-n_{\kappa,2}^*=\left(R_{\kappa,2}^{*}\right)^{1/2}\nu_{\kappa,2},
\label{eq:16x}
\end{eqnarray}
at each saddle point $\kappa=a,b$, where $R_{\kappa,i}^{*}$ are the reaction rates at the two saddle points.  Then the free energy Eq.~(\ref{eq:12x}) takes the form
\begin{equation}
W\left(n_1,n_2\right)-W_{\kappa}^{*} = \frac{1}{2}\sum_{i,j}\nu_{\kappa,i} \Gamma_{\kappa,ij}^{*} \nu_{\kappa,j},\;\;\;\kappa=a,b
\label{eq:17x}
\end{equation}
where
\begin{equation}
\Gamma_{\kappa,ij}^{*}=\left(R_{\kappa,i}^{*}\right)^{1/2}W_{\kappa,ij}^{*}\left(R_{\kappa,j}^{*}\right)^{1/2}
\label{eq:18x}
\end{equation}
includes the information of not only the free energy landscape $W$ but also the reaction rates $R$.  By diagonalizing Eq.~(\ref{eq:18x}) as
\begin{equation}
W-W_{\kappa}^{*}=\frac{1}{2}\left(\lambda_\kappa\xi_\kappa^{2}+\gamma_\kappa\eta_\kappa^{2}\right)
\label{eq:19x}
\end{equation}
using the orthogonal transformation~\cite{Wilemski1999}
\begin{eqnarray}
\xi_\kappa &=& \nu_{\kappa, 1}\cos\alpha_{\kappa}+\nu_{\kappa, 2}\sin\alpha_{\kappa}
\label{eq:20x} \\
\eta_\kappa &=& -\nu_{\kappa, 1}\sin\alpha_{\kappa}+\nu_{\kappa, 2}\cos\alpha_{\kappa}
\label{eq:21x}
\end{eqnarray}
with
\begin{equation}
\tan\alpha_{\kappa}=\left(\Gamma_{\kappa, 11}-\Gamma_{\kappa, 22}-G_{\kappa}\right)/\left(2\Gamma_{\kappa, 12}\right)
\label{eq:22x}
\end{equation}
and
\begin{equation}
G_{\kappa}=\sqrt{\left(\Gamma_{\kappa, 11}-\Gamma_{\kappa, 22}\right)^{2}+4\left(\Gamma_{\kappa, 12}\right)^{2}},
\label{eq:23x}
\end{equation}
we can obtain the eigenvalues $\lambda_{\kappa}$ and $\gamma_{\kappa}$ given by
\begin{eqnarray}
\lambda_{\kappa}&=&\left(\Gamma_{\kappa, 11}+\Gamma_{\kappa, 22}-G_{\kappa}\right)/2<0,
\label{eq:24x} \\
\gamma_{\kappa}&=&\left(\Gamma_{\kappa, 11}+\Gamma_{\kappa, 22}+G_{\kappa}\right)/2>0.
\label{eq:25x}
\end{eqnarray}
Then, the magnitude of the steady-state nucleation rate $J_{\kappa}^{\rm st}$ at the saddle point $\kappa=a, b$ is given simply by
\begin{equation}
J_{\kappa}^{\rm st}=N_{\kappa}^{*}\sqrt{\frac{R_{\kappa 1}^{*}R_{\kappa 2}^{*}\left|\lambda_{\kappa}\right|}{\gamma_{\kappa}}}
\label{eq:26x}
\end{equation}
using the eigenvalues $\lambda_{\kappa}$ and $\gamma_{\kappa}$~\cite{Trinkaus1983,Wilemski1999}.
The total nucleation rate can be calculated by Eqs.~(\ref{eq:10x}) and (\ref{eq:26x}).  Therefore, the total nucleation rate will be larger than the individual nucleation rates $J_{a}^{\rm st}$ and $J_{b}^{\rm st}$ as the extra channel of nucleation reaction exists. 

Since $N_{\kappa}^{*}=N_{0}\exp\left(-W_{\kappa}^{*}/kT\right)$, the two fluxes $J_{a}^{\rm st}$ and $J_{b}^{\rm st}$ will contribute almost equally when the free-energy barriers $W_{\kappa}^{*}$ are almost the same magnitude.  An example of the free-energy landscape when two saddle points have the same free-energy barriers has already be shown in Fig.~\ref{fig:1x}.

Even when two competitive routes $\kappa=a$ and $\kappa=b$ do not have the same free energy barriers, the corresponding two competitive nuclei nucleate probabilistically.  A qualitatively similar conclusion was reached by Sanders et al.~\cite{Sanders2007} using the artificially designed {\it q}-states Potts model.  The probability is in fact proportional to the Boltzmann factor  $N_{\kappa}^{*}\propto\exp\left(-W_{\kappa}^{*}/kT\right)$ from Eq.~(\ref{eq:26x}).

\subsection{\label{sec:level3}Series nucleation}

\begin{figure}
\includegraphics[width=0.65\linewidth]{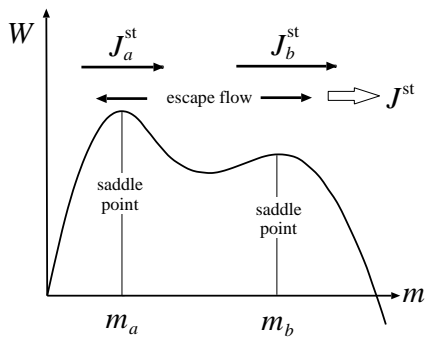}
\caption{Free-energy landscape $W$ when there exist two saddle points $m_{a}$ and $m_{b}$ along the fictitious one-dimensional axis $m$ shown schematically in Fig.~\ref{fig:2x}(b). }
\label{fig:3x}
\end{figure}

When two saddle points $a$ and $b$ appear sequentially along the line which is almost perpendicular to the line $n_1+n_2\sim\mbox{constant}$, these two saddle points corresponds to the two types of nucleus of almost the same composition $n_2\propto n_1$ with different sizes (small and large).  In such a case, it is reasonable to assume that the nucleation flux ${\bm J}$ will remain one stream as shown schematically in Fig.~\ref{fig:2x}(b) and the problem will be reduced to the one-dimensional double-barrier~\cite{Valencia2004} problem shown in Fig.~\ref{fig:3x}.

By changing the coordinate system along the flux, the problem can be mapped to the one dimensional problem with two barriers first studied by Valencia and Lipowsky~\cite{Valencia2004}.  Suppose we map the nucleation flux on the two-dimensional coordinates system $(n_1,n_2)$ to one dimensional system with coordinate $m$ (Fig.~\ref{fig:2x}(b)), then the nucleation flux Eq.~(\ref{eq:6x}) becomes one-dimensional
\begin{equation}
J=-N\left(m\right)R_{\rm eff}\left(m\right)\frac{d}{d m}\left(\frac{f\left(m\right)}{N\left(m\right)}\right),
\label{eq:27x}
\end{equation}
where we have used the definition of the potential $\Phi$ in Eq.~(\ref{eq:4x}) and introduced an effective reaction rate $R_{\rm eff}\left(m\right)$.  In the steady state, the nucleation flux is constant $J=J^{\rm st}$ and we have
\begin{eqnarray}
J^{\rm st}\int_{0}^{\infty}\frac{dm}{R_{\rm eff}\left(m\right)N\left(m\right)}=-\int_{0}^{\infty}\frac{d}{dm}\left(\frac{f\left(m\right)}{N\left(m\right)}\right)
\nonumber \\
=\frac{f\left(m\rightarrow 0\right)}{N\left(m\rightarrow 0\right)}-\frac{f\left(m\rightarrow\infty\right)}{N\left(m\rightarrow\infty\right)}.
\label{eq:28x}
\end{eqnarray}
from Eq.~(\ref{eq:27x}).  Using the boundary condition~\cite{McDonald1963} in Eq.~(\ref{eq:8x}), we found
\begin{equation}
J^{\rm st}=\left[ \int_{0}^{\infty}\frac{dm}{R_{\rm eff}\left(m\right)N\left(m\right)}\right]^{-1}.
\label{eq:29x}
\end{equation}
By dividing the integrals into two parts around the two saddle point $m_{a}$ and $m_{b}$ as
\begin{equation}
J^{\rm st}=\left[ \int_{\left(m_{a}\right)}\frac{dm}{R_{\rm eff}\left(m\right)N\left(m\right)}
+\int_{(m_{b})}\frac{dm}{R_{\rm eff}\left(m\right)N\left(m\right)}
\right]^{-1},
\label{eq:30x}
\end{equation}
we arrive at the formula
\begin{equation}
\frac{1}{J^{\rm st}}=\frac{1}{J_{a}^{\rm st}}+\frac{1}{J_{b}^{\rm st}}
\label{eq:31x}
\end{equation}
for the series nucleation that is similar to the formula for the conductivity of two resistors connected in series, where
\begin{equation}
J_{\kappa}^{\rm st}=R_{\rm eff}\left(m_\kappa\right)N\left(m_\kappa\right)Z_\kappa,\;\;\;\kappa=a,b
\label{eq:32x}
\end{equation}
with the Zeldovich factor
\begin{equation}
Z_\kappa=\sqrt{\left|\partial^{2}W\left(m_\kappa\right)/\partial m^{2}\right|/2\pi kT}
\label{eq:33x}
\end{equation}
is the nucleation rate when the saddle point $\kappa=a, b$ exists independently.  The factor $1/\sqrt{2\pi kT}$ appears in Eq.~(\ref{eq:33x}) of one-dimensional problem in contrast to Eq.~(\ref{eq:26x}) of two-dimensional problem since the general formula for the nucleation rate by Trinkaus~\cite{Trinkaus1983} for the $d$-dimensional free-energy surface is proportional to a factor $\left(1/2\pi kT\right)^{1-d/2}$.  Eq.~(\ref{eq:31x}) was initially derived by Valencia and Lipowsky~\cite{Valencia2004} using the Master equation and later re-derived by Valencia~\cite{Valencia2006} using the Kramers theory.   
Here, we have directly used the continuum description and the Fokker-Planck equation.  Reduction of the nucleation rate $J^{\rm s}$ from an independent $J_\kappa^{s}$ is due to the escape flow from intermediate well (Fig.~\ref{fig:3x})~\cite{Valencia2006}.  The double-barrier problem was also studied using Kramers theory by Nicolis and Nicolis~\cite{Nicolis2003}.  Since they did not use the standard theory of nucleation based on the Fokker-Planck or the Zeldovich-Frenkel equation, they could not derive a simple expression for the total nucleation rate like Eq.~(\ref{eq:31x}).

\section{\label{sec:level3x}Concluding remark}
\begin{figure}
\includegraphics[width=0.65\linewidth]{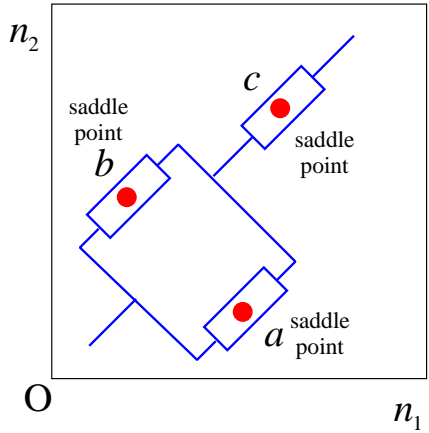}
\caption{Complex nucleation process represented by an analogy to the electric circuit when there exist multiple saddle points $a$,  $b$ and $c$ on the free-energy landscape $W$. Total steady-state nucleation rate $J^{\rm st}$ is given by $\frac{1}{J^{\rm st}}=\frac{1}{J_a^{\rm st}+J_b^{\rm st}}+\frac{1}{J_c^{\rm st}}$. }
\label{fig:4x} 
\end{figure}

In this report, we have summarized the basic ingredients of parallel and series nucleation when there exist two saddle points on the free-energy landscape.  Extension to the more complex nucleation process is possible by drawing the diagram similar to the electric circuitry as shown in Fig.~\ref{fig:4x}.  Figure 4 shows an example of complex nucleation when there are three saddle points on the free energy landscape.  In this case, the steady-state nucleation rate will be calculated from
\begin{equation}
\frac{1}{J^{\rm st}}=\frac{1}{J_a^{\rm st}+J_b^{\rm st}}+\frac{1}{J_c^{\rm st}}.
\label{eq:34x}
\end{equation}
Extension to more complex situation is obvious.

However, it must be noted, our analysis is completely confined to the steady-state process.  Therefore, transient properties are out of our scope, which can only be studied numerically by solving coupled Master equation~\cite{Wilemski1995,Wyslouzil1996}.  Also, our analysis assumed that the nucleation flux goes through the saddle point.  Saddle point avoidance~\cite{Trinkaus1983,Wyslouzil1995} will be important if the anisotropy of the reaction matrix $R$ is large or the ridge between the saddle points is law, which can occur at high temperatures or near the spinodal point.  In such a case, ridge-crossing rather than saddle-crossing may occur.  Then the nucleation flux will spread over the whole phase space and our description may break down.

\begin{acknowledgments}
This work was supported by the Grant-in-Aid for Scientific Research [Contract No.(C)22540422] from Japan Society for the Promotion of Science (JSPS)  and a project for strategic advancement of research infrastructure for private universities, 2009-2013, from MEXT, Japan. 
\end{acknowledgments}

\nocite{*}

\begin{thebibliography}{99}
\bibitem{Kelton2010} K. F. Kelton and A. L. Greer, Nucleation in Condensed Matter, Applications in Materials and Biology, Pergamon, Oxford, 2010, Chapter 6.
\bibitem{Vekilov2004} P. G. Vekilov, Cryst. Growth. Des. {\bf 4}, 671 (2004).
\bibitem{Gebauer2008} D. Gebauer, A. V\"olkel, and H. G\"olfen, Science {\bf 322}, 1819 (2008).
\bibitem{Ray1986} A. K. Ray, M. Chalam, and L. K. Peters, J. Chem. Phys. {\bf 85}, 2161 (1986).
\bibitem{Valencia2004} A. Valencia and R. Lipowsky, Langmuir {\bf 20}, 1986 (2004).
\bibitem{Kashchiev1998} D. Kashchiev and K. Sato, J. Chem. Phys. {\bf 109}, 8530 (1998).
\bibitem{Kashchiev2005} D. Kashchiev, P. G. Vekilov, and A. B. Kolomeisky, J. Chem. Phys. {\bf 122}, 244706 (2005).
\bibitem{Iwamatsu2011} M. Iwamatsu, J. Chem. Phys. {\bf 134}, 164508 (2011).
\bibitem{Chen2003} B. Chen, J. I. Siepmann, and M. L. Klein, J. Am. Chem. Soc. {\bf 125}, 3113 (2003).
\bibitem{Wagner2001} P. E. Wagner and R. Strey, J. Phys. Chem. B {\bf 105}, 11656 (2001).
\bibitem{Wu1997} D. T. Wu, Sol. St. Phys. {\bf 50}, 38 (1997).
\bibitem{Risken1989} H. Risken: The Fokker-Planck Equation, 2nd ed, Springer, Berlin 1989.
\bibitem{Zeldovich1942} J. Zeldovich, Sov. J. Exp. Theor. Phys. {\bf 12}, 525 (1942). 
\bibitem{Frenkel1955} J. Frenkel: Kinetic Theory of Liquids, Dover, New York 1955.
\bibitem{Reiss1950} H. Reiss, J. Chem. Phys. {\bf 18}, 840 (1950).
\bibitem{Temkin1984} D. E. Temkin and V. V. Shevelev, J. Cryst. Growth {\bf 66}, 380 (1984).
\bibitem{Stauffer1976} D. Stauffer, J. Aerosol Sci. {\bf 7}, 319 (1976).
\bibitem{Trinkaus1983} H. Trinkaus, Phys. Rev. B {\bf 27}, 7372 (1983).
\bibitem{Greer1990} A. L. Greer, P. V. Evans, R. G. Hamerton, D. K. Shangguan and K. F. Kelton, J. Cryst. Growth. {\bf 99}, 38 (1990).
\bibitem{Wilemski1995} G. Wilemski and B. Wyslouzil, J. Chem. Phys. {\bf 103}, 1127 (1995)
\bibitem{Wyslouzil1995} B, Wyslouzil and G. Wilemski, J. Chem. Phys. {\bf 103}, 1137 (1995)  
\bibitem{Wilemski1999} G. Wilemski, J. Chem. Phys. {\bf 110}, 6451 (1999).
\bibitem{Fisenko2004} S. P. Fisenko and G. Wilemski, Phys. Rev. E {\bf 70}, 056119 (2004).
\bibitem{Sanders2007} D. P. Sanders, H. Larralde, and F. Leyvaz, Phys. Rev. B {\bf 75}, 132101 (2007).
\bibitem{McDonald1963} J. E. McDonald, Am. J. Phys. {\bf 31}, 31 (1963).
\bibitem{Valencia2006} A. Valencia, J. Chem. Phys. {\bf 125}, 144704 (2006).
\bibitem{Nicolis2003} G. Nicolis and C. Nicolis, Physica A {\bf 323}, 139 (2003).
\bibitem{Wyslouzil1996} B. Wyslouzil and G. Wilemski, J. Chem. Phys. {\bf 105}, 1090 (1996).



\end{thebibliography}

\end{document}